\documentclass[aps,prb,twocolumn,showpacs,showkeys,groupedaddress,floatfix,amsmath,amssymb]{revtex4}

\usepackage{graphicx}
\usepackage{dcolumn}
\usepackage{bm}

\begin{document}

\title{Stable and metastable states in a superconducting "eight" loop in applied magnetic field}

\author{D. Y. Vodolazov}
\author{F. M. Peeters}
\email{peeters@uia.ua.ac.be} \affiliation{Departement Natuurkunde,
Universiteit Antwerpen (UIA), Universiteitsplein 1,  B-2610
Antwerpen, Belgium}

%\date{\today}

\begin{abstract}

The stable and metastable states of different configurations of a
loop in the form of an eight is studied in the presence of a
magnetic field. We find that for certain configurations the
current is equal to zero for any value of the magnetic field
leading to a magnetic field independent superconducting state. The
state with fixed phase circulation becomes unstable when the
momentum of the superconducting electrons reaches a critical
value. At this moment the kinetic energy of the superconducting
condensate becomes of the same order as the potential energy of
the Cooper pairs and it leads to an instability. Numerical
analysis of the time-dependent Ginzburg-Landau equations shows
that the absolute value of the order parameter changes gradually
at the transition from a state with one phase circulation to
another although the vorticity change occurs abruptly.

\end{abstract}


\maketitle

Starting with the pioneering work of Little and Parks
\cite{Little} who studied the properties of double-connected
superconductors (confined samples with two surfaces, for example
rings) in applied magnetic field there has been continued
theoretical and experimental interest in those systems. Changing
the topology of the system is expected to have a strong influence
on the superconducting properties, in particular in the presence
of an external magnetic field. For instance in Refs.
\cite{Mila,Hayashi} the M\"obuis loop was considered and it was
obtained that the phase diagram of this system differs
considerably from an ordinary loop \cite{Hayashi}.

In the present paper we investigate a different topology, a loop
having an "eight" geometry. The latter can be considered as a
combination of two strongly interacting loops (see Fig.1). In
comparison with two unconnected loops the direction and value of
the current density in one loop of the "eight" system will depend
on the direction and value of the current in the other loop. As a
result, the free energy of such a system will not be merely the
sum of the free energy of two unconnected loops. Moreover, we
found that for certain values of the relative sizes of the two
loops the current density in such a loop will be equal to zero at
any value of the applied magnetic field.

In this work we study the stable and metastable states of a loop
in the shape of an "eight". This geometry can be obtained by
grapping two radial sides of a single loop and twisting it over
180 degrees with respect to each other. We assume that there is no
physical contact at the crossing point of the "eight". We can
further distinguish three configurations for this system as shown
in Fig. 2. The "eight" loop was modelled as a combination of two
rings with radius $R_1$ and $R_2$. We use the Ginzburg-Landau (GL)
theory in order to study the structure of the superconducting
state of such a system.
\begin{figure}[h]
\includegraphics[width=0.58\textwidth]{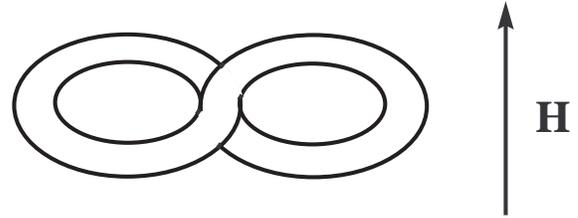}
\caption{The configuration.}
\end{figure}
We neglect the finite width $w$ and thickness $d$ of the loop.
This is allowed when the radius of the loop is much larger than
$w$ and if $d<<\lambda$, where $\lambda$ is the London penetration
length. The last condition also allows us to neglect the screening
effects (self-inductance of the system). Within these
simplifications the GL equations are one-dimensional and using the
gauge-invariant momentum $p$ we can write it in the following form
\begin{subequations}
\begin{eqnarray}
 \frac{d^2f}{ds^2}+f(1-f^2-p^2)=0,  \label{a}%1a
\\
 j=f^2p.  \label{b} %1b
\end{eqnarray}
\end{subequations}
where $\psi=f(s)e^{{\rm i}\phi(s)}$ is a undimensional order
parameter, the momentum $p=\nabla \phi -A$ is scaled in units
$\Phi_0/(2\pi\xi)$ (where $\Phi_0$ is the quantum of magnetic
flux), the length of the loop is $L=2\pi(R_1+R_2)$ and the arc
coordinate $s$ is in units of the coherence length $\xi(T)$. In
these units the magnetic field is scaled in the unit $H_{c2}$ and
the current density $j$ in $j_0=c\Phi_0/8\pi^2\lambda^2\xi$.

\begin{figure}[h]
\includegraphics[width=0.5\textwidth]{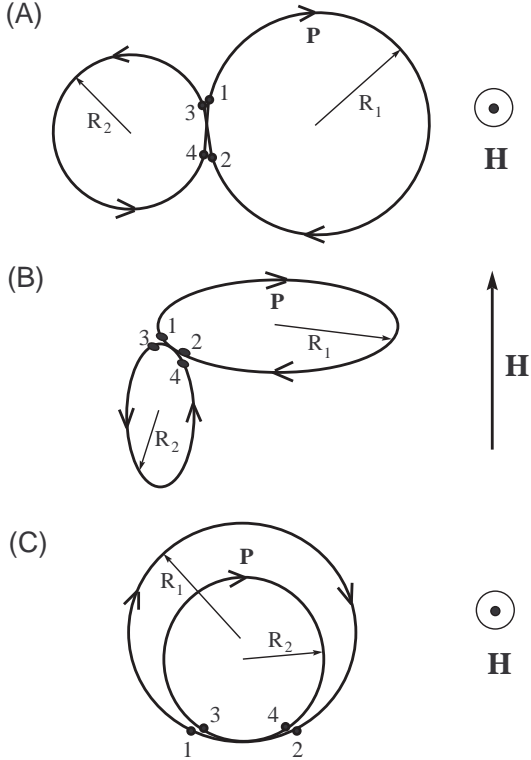}
\caption{Different 1D model configurations for the "eight" loop.}
\end{figure}

At first we find the dependence of the momentum of the "eight"
loop on the applied magnetic field. From equation ${\rm div} {\bf
j}=0$ it follows that the supercurrent $j=f^2p$ is constant all
over the system. Our numerical analysis of Eqs. (1a,b) shows that
in the stationary state, $F$ does not depend on $s$. As a result
$p=const$ along the loop and the order parameter is equal to
$f=\sqrt{1-p^2}$.

Let us first consider the circulation of the momentum $p$ for
system $(A)$ of Fig. 2
\begin{equation}
\oint {\bf p} \cdot {\bf ds}=2\pi(R_1+R_2)p=2\pi n-\int_1^2{\bf %2
A} \cdot {\bf ds}- \int_3^4{\bf A} \cdot {\bf ds},
\end{equation}
where the points 1,2,3,4, are at the twist point of the loop which
in Fig. 2 are shown slightly separated from this point for
clarity. As a result we find
\begin{equation}
p=\frac{1}{R_1+R_2}(n-(\Phi_1-\Phi_2)), %3
\end{equation}
where $\Phi_{1,2}=HR_{1,2}^2/2$ is the magnetic flux through the
rings with radius $R_1$ and $R_2$, respectively (in $\Phi_0$
units), $n=\oint \nabla \phi ds/2\pi$ is an integer number
defining the circulation (or vorticity) of the phase of the order
parameter in the ring . From Eq. (3) follows the interesting
property that for $R_1=R_2$ the momentum and hence the current
density will be equal to zero for any value of the applied
magnetic field in system $(A)$. Thus this system will have no
response to an applied magnetic field and consequently
superconductivity should be conserved up to very high magnetic
fields. The other consequence of it is that there is no
oscillations in the phase diagram $H-T$ of such a system.

It is easy to show that for the systems $(B,C)$ the dependence
$p(H)$ are, respectively, given by
\begin{subequations}
\begin{eqnarray}
p=\frac{1}{R_1+R_2}(n-\Phi_1), \label{a}%4a
\\
p=\frac{1}{R_1+R_2}(n-(\Phi_1+\Phi_2)). \label{b}%4b
\end{eqnarray}
\end{subequations} In the limit $R_2 \to 0$ we obtain from Eqs.
(3,4) the well known result
\begin{equation}
p=\frac{1}{R_1}(n-\Phi_1), %5
\end{equation}
which is valid for a single double connected loop (see for example
Ref. \cite{Tinkham}). It is interesting to note that for system
$(B)$ the value for $p$ is a factor $(R_1/(R_1+R_2))$ less as
compared to the single loop case.

Using Eqs. (3,4) we can find the stable states of our systems
$(A-C)$. This is determined by the global minimum of the Gibbs
free energy
\begin{equation}
G[f]=\oint
\left(\left(\frac{df}{ds}\right)^2-f^2+\frac{f^4}{2}+f^2p^2 %6
\right)ds,
\end{equation}
resulting in a minimum value of $p$ at given $H$. The dependence
of $p(H)$ is a periodic function of $H$ with period
$2/(R_1^2-R_2^2)$, $2/R_1^2$ and $2/(R_1^2+R_2^2)$ for systems
$(A)$, $(B)$, $(C)$, respectively. Note that the maximum value of
$p_{max}$ for the thermodynamically stable state of our systems
$(A-C)$ is the same:  $1/2(R_1+R_2)$.

\begin{figure}[h]
\includegraphics[width=0.5\textwidth]{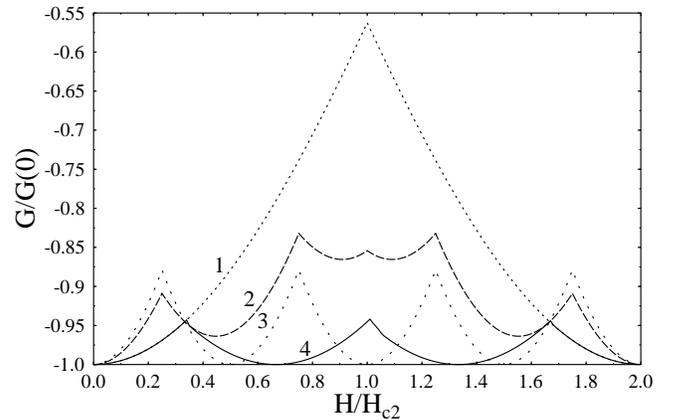}
\caption{The Gibbs free energy of the "eight" loop in the $(A)$
configuration (curve 4) with parameters $R_1=2\xi$, $R_2=\xi$ and
the energies of single rings (with radius $R=2\xi$ (curve 3) and
$R=\xi$ (curve 1)). Curve 2 corresponds to the sum of the energy
of these two uncoupled rings.}
\end{figure}

The two rings in our "eight" loop are strongly coupled and as a
result the Gibbs free energy of this system is not simply the sum
of the energies of two uncoupled rings plus a small interaction
term as for the case of two magnetically coupled rings
\cite{Baelus,Morelle}. In Fig. 3 the dependence of the equilibrium
energy of the "eight" system (case $(A)$ - curve 4) and two
separate rings with different radii (curves 1, 3) are presented.
In the same figure the sum of the Gibbs free energies of the two
uncoupled rings is shown (curve 2). The dependence $G(H)$ of the
"eight" loop is more similar to the dependency of $G(H)$ for the
separate rings then with the sum of the two energies.

Analysis shows that the state of the system for a specific value
of the phase circulation $n$ may be stable (more exactly
metastable) for $|p|>p_{max}$. This state becomes unstable when
the second variation of the Gibbs free energy (6) becomes equal to
zero. Let's consider small deviations $f_{\epsilon}\ll f$,
$p_{\epsilon}\ll p $ (where the full solution are $\tilde
f=f+f_{\epsilon}$ and $\tilde p=p+p_{\epsilon}$) from the stable
solutions of our systems \cite{Fink,Kramer}. The Ginzburg-Landau
equations linearized with respect to the small perturbations $f$
and $p$ are given by
\begin{subequations}
\begin{eqnarray}
 \frac{d^2f_{\epsilon}}{ds^2}+f_{\epsilon}(1-3f^2-p^2)-2fpp_{\epsilon}=0, \label{a}%7a
\\
f^2p_{\epsilon}+2fpf_{\epsilon}=C. \label{b}%7b
\end{eqnarray}
\end{subequations}
The constant $C$ in Eq. (7a) has to be equal to zero because
otherwise $p_{\epsilon} \neq 0$ and $f_{\epsilon} \neq 0$ for any
$f$ and $p$. Using Eq. (7b) we solve for $p_{\epsilon}$ and
substitute it in Eq. (7a) in order to obtain the following
equation
\begin{equation}
\frac{d^2f_{\epsilon}}{ds^2}+f_{\epsilon}(6p^2-2)=0, %8
\end{equation}
which has the following solution
\begin{equation}
f_{\epsilon}(s)=A{\rm cos}(\omega s)+B{\rm sin}(\omega s), %9
\end{equation}
with $\omega=\sqrt{6p^2-2}$. Taking into account the boundary
condition $f_{\epsilon}(0)=f_{\epsilon}(L)$ we find that a state
with fixed $n$ becomes unstable when the following condition is
fulfilled
\begin{equation}
p=p_c=\frac{1}{\sqrt{3}}\sqrt{1+\frac{1}{2(R_1+R_2)^2}} \, . %10
\end{equation}
Inserting $R_2=0$ in Eq. (10) we obtain the stability condition of
a single loop as was obtained in Ref. \cite{Horane} (expressed in
$\Phi$, $n$ and $\xi/R_1$). In the case of large (i.e. $ {\rm
min}[R_1,R_2] \gg \xi$) loops $p_c=1/\sqrt{3}$ and the current
density is equal to $j=p_c(1-p_c^2)=j_c=2/3\sqrt{3}$ - which is
the depairing current density. Thus Eq. (10) has a simple physical
meaning: when the kinetic energy of the superconducting condensate
(or Cooper pair) becomes of the order of the potential energy the
state becomes unstable. The finite length of the loop starts to
play an essential role when $L$ is about the Cooper pair size
which results in different values for $p_c$.
\begin{figure}[h]
\includegraphics[width=0.48\textwidth]{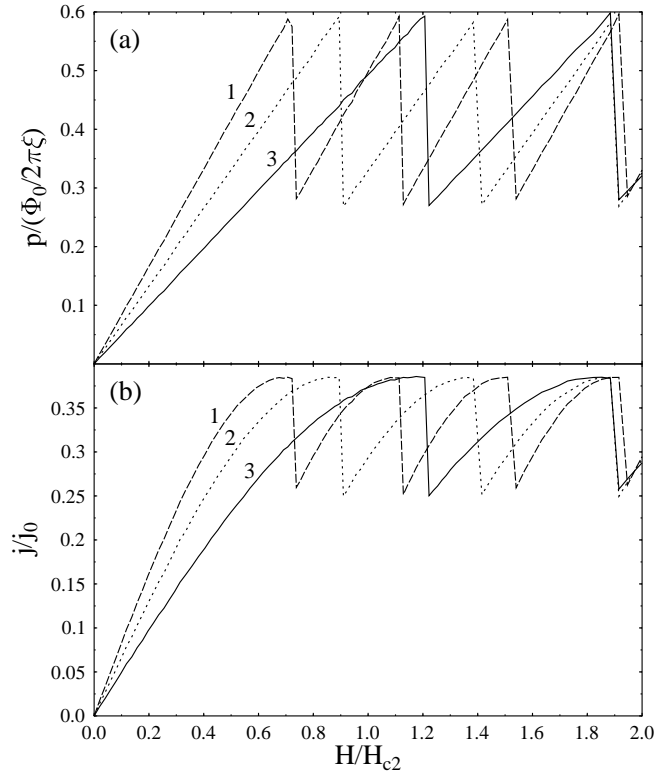}
\caption{Magnetic dependencies of $p(H)$ (a) and $j(H)$ (b) for
the three systems shown in Fig. 2. Curves 1,2,3 correspond to the
systems $(A)$, $(B)$, $(C)$, respectively, with $R_1=2\xi$ and
$R_2=\xi$.}
\end{figure}

Using Eq. (10) it is easy to find the critical values of the
magnetic field $H_{c,n}$ at which the transition from state $n$ to
state $n+1$ occurs. Consider for example system $(A)$ (see Fig.
2), and
\begin{figure}[hbtp]
\includegraphics[width=0.48\textwidth]{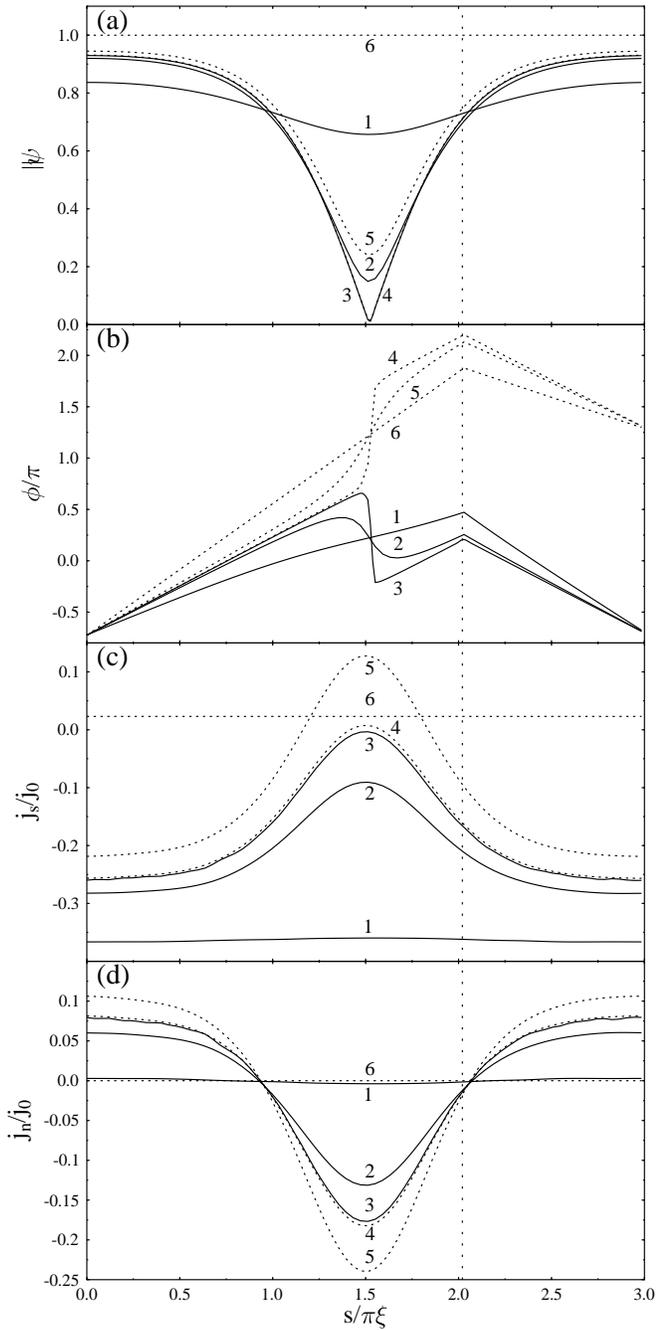}
\caption{Dependencies of the order parameter (a), the phase of
order parameter (b), the superconducting (c) and normal (d)
current density when switching the applied magnetic field from
$2.51H_{c2}$ to $2.52 H_{c2}$. The interval $(0,2)$ corresponds to
the ring with radius $R_1=\xi$ and $(2,3)$ to the ring with radius
$R_2=0.5\xi$. Curves 1, 2, 3, 4, 5, 6 are at the times 500, 512.5,
512.95, 513, 513.5, 550 $\tau$ respectively. Curves 1, 2, 3,
correspond to times less than $t_0$ which is the time at which the
order parameter equals zero at one point on the ring (for our
parameters $t_0 \simeq 512.98 \tau$).}
\end{figure}
combine Eq. (10) with Eq. (3) we obtain
\begin{equation}
H_{c,n}=\frac{2}{R_1-R_2}\left[\frac{1}{\sqrt{3}}
\sqrt{1+\frac{1}{2(R_1+R_2)^2}}-\frac{n}{R_1+R_2} \right]. %11
\end{equation}
It is interesting that for the configurations $(B)$ and $(C)$ the
critical fields will be different but $p_c$ (and $p_{max}$) will
be the same under the condition that the length of the loops are
the same. Moreover for loops with $L/\xi \gg 1$ the value of the
critical momentum does not depend on the size of the system.
Therefore, we may conclude that condition (10) is an universal
condition for one-dimensional double-connected systems of the type
shown in Fig. 2. Even if the loops have a different shape, for
example ellipsoid, the condition for vortex entry will be
described by Eq. (10) (with the change of $(R_1+R_2)$ by $L/2\pi$)
because the order parameter in such a sample will also be uniform
along the loop. Indeed, a loop of arbitrary shape with length $L$
is equivalent to a one-dimensional wire of length $L$ with
periodical boundary conditions. The presence of a magnetic field
leads to a nonuniform (uniform for a single circular ring)
distribution of the vector potential along the wire which is
compensated by the term $\nabla \phi$ in such a way that the
current density and the order parameter are uniform along the wire
\cite{comm1}.

We checked Eq. (10) through a numerical solution of the
time-dependent one-dimensional Ginzburg-Landau equations
\begin{subequations}
\begin{eqnarray}
 -\gamma \left(\frac {\partial \psi}{\partial t}+i\varphi\psi  % 12a
\right)=(-{\rm i}\nabla - {\bf A})^2 \psi +\psi(|\psi|^2-1),
\qquad
\\
\frac{d}{ds}\left[ \frac{d\varphi}{ds}-{\rm Re}(\psi^*(-{\rm
i}\nabla-{\bf A})\psi)\right]=0.
\qquad % 12b
\end{eqnarray}
\end{subequations}
Here time is scaled in units of $\tau=4\pi\sigma_n \lambda^2
(T)/c^2 $, the electrostatic potential $\varphi$ in units of
$c\Phi_0/8 \pi^2 \xi \lambda \sigma_n$ ($\sigma_n $ is the
normal-state conductivity), and $\gamma$ is a relaxation constant
\cite{Ivlev}.

In Fig. 4 the dependency of $p$ (Fig. 4(a)) and the current
density (Fig. 4(b)) on the applied magnetic field are shown for
the three systems (we took $R_1=2\xi$, $R_2=1 \xi$). The magnetic
field was increased linearly from $H=0$ to $2H_{c2}$ during a time
period of $\Delta t=2 \cdot 10^4\tau$. The theoretical value of
$p_c$ for such a system is $p_c=0.593$. From our numerical
calculation we obtained $p_c \simeq 0.596$. The difference between
the numerical result and Eq. (10) is in the range of our numerical
error.

From our numerical solution of the time-dependent Ginzburg-Landau
equations we find the evolution of the system from a state with
phase circulation $2\pi n$ to a state with $2\pi (n+1)$. In Fig. 5
the dependence of $|\psi|$, $\phi$, $j_s$ and $j_n$ are shown for
the system $(A)$ with parameters $R_1=\xi$, $R_2=0.5\xi$. The
magnetic field was increased gradually to $2.51H_{c2}$ and then in
one step to $2.52 H_{c2}$. At time $t=0$ the "eight" loop was at a
metastable state at $H=2.51H_{c2}$. At $t=t_0\simeq 513 \tau$
(curve 4 in Fig. 5) the order parameter reaches zero at a certain
point on the ring. Simultaneously the phase difference $\Delta
\phi=\phi(+0)-\phi(-0)$ near the zero point becomes equal to
$-\pi$. In Refs. \cite{Langer,Vodol1} the vanishing of the order
parameter in one point on a single ring was considered and it was
found that when such a solution exists the superconducting current
density is equal to zero and $\Delta \phi = \pm \pi$ (sign depends
on the current direction). Although in our time-dependent problem
the full current is not equal to zero, nevertheless there is a
similarity between the stationary and the non-stationary problems
- in both cases when the order parameter reaches zero there is a
phase shift of $\Delta \phi=\pm \pi$ near the point where
$\psi=0$.

For $t>t_0$ the superconducting current (see Fig. 5(c)) changes
sign near the minimum point and hence it leads to a phase change
of $\Delta \phi=2\pi$ (from $-\pi$ to $+\pi$) according to
\cite{Vodol1} (see also Fig. 5(b)). As a result a phase
circulation of $\oint \nabla \phi ds$=$2\pi$ appears in the ring (
see Fig. 5(b)). Then the order parameter increases in the point
$\psi_{min}$ and $\Delta\phi$ gradually decreases to 0 (see Fig.
5(b)). This is also in qualitative agreement with the results of
Refs. \cite{Langer,Vodol1} where it was found that for increasing
$|\psi|_{min}$ (or $|j_s|$) $\Delta \phi$ decreased to zero near
the minimum point.

Because $\nabla \phi$ is a continuous function everywhere in the
loop, the current density (superconducting $j_s$ and normal $j_n$)
will also change continuously. The time evolution of $j_s$ and
$j_n$ are shown in Fig. 5(c) and Fig. 5(d), respectively. These
currents are a function of the arc-coordinate during the
transition process and only the sum $j_s+j_n$ does not depend on
$s$. For large times the normal current density decays and the
order parameter becomes uniform in the system.

In conclusion, the stable and metastable states of three different
configurations for the "eight" loop in a magnetic field were
studied. We found that the state with fixed phase circulation
becomes unstable when the value of the momentum of the
superconducting electrons reaches a critical $p_c$. Then the
kinetic energy of the superconducting condensate becomes of the
same order as the potential energy of the Cooper pairs resulting
in an instability. Numerical analysis of the time-dependent
Ginzburg-Landau equations shows that the absolute value of the
order parameter changes gradually at the transition from a state
with one phase circulation to another although the vorticity
changes abruptly.

The work was supported by the Flemish Science Foundation (FWO-Vl),
the "Onderzoeksraad van de Universiteit Antwerpen," the
"Interuniversity Poles of Attraction Program - Belgian State,
Prime Minister's Office - Federal Office for Scientific, Technical
and Cultural Affairs," and the European ESF-Vortex Matter. One of
us (D.Y.V.) is supported by a postdoctoral fellowship of FWO-Vl.


\begin{references}
\bibitem{comm1} This statement is valid only for loops which are connected
as shown in Fig. 2. For systems in which there are points with
three and more connected wires, the distribution of the order
parameter will be nonuniform along the system (see \cite{Fink1})
and criterion (10) is not valid.
\end{references}
\end{document}